# Project SPARROW and the Future of Conservation Technology

Juan M. Lavista Ferres*[,1], Carl Chalmers*[,1], Bruno Demuro Segundo*[,1], Zhongqi Miao*[,1], Andres Hernandez Celis[1,2], Federico Alves Torres[1], Isai Daniel Chacon Silva[1,2], Anthony Cintron Roman[1], Allen Kim[1], Meygha Machado[1], Luana Marotti[1], Amy Michaels[1], Daniela Ruiz Lopez[1,2], Catherine Romero[1], Rahul Dodhia[1], Inbal Becker-Reshef[1,3], Pablo Arbelaez[2]

*These authors contributed equally to this work.

[1]Microsoft AI for Good Lab, Redmond.

[2]Universidad de los Andes, Bogotá.

[3]University of Maryland, College Park.

## Abstract

Global biodiversity is declining at unprecedented rates, yet the tools available to monitor and protect ecosystems remain limited by constraints in power, connectivity, and accessibility. We present SPARROW, a hardware and software open-source platform that integrates solar energy, edge artificial intelligence, and satellite communication to enable continuous, autonomous biodiversity monitoring in remote environments. Each SPARROW node combines a low-power Graphics Processing Unit (GPU) with modular visual, acoustic, and environmental sensors, performing on-device deep learning inference and transmitting summarized results through Low-Earth-Orbit (LEO) satellite or Global System for Mobile Communications (GSM) networks.

We deployed SPARROW across tropical, temperate, and montane ecosystems in Colombia, Peru, Tanzania, and the United States, where it sustained 24/7 operation under variable environmental conditions and collected more than two million images and acoustic recordings in the first 190 days. The system demonstrated robust real-time classification and adaptive power management, achieving full autonomy without on-site human intervention. By integrating renewable energy, on-edge AI, and open-source design, SPARROW lowers the technical and financial barriers to ecological monitoring and establishes a scalable foundation for a distributed, intelligent network of sensors, an emerging “Internet of Living Things” for planetary biodiversity monitoring.

## 1. Introduction

Over the past fifty years, we have witnessed a troubling erosion of our planet’s biodiversity. According to the WWF’s Living Planet Report, populations of monitored vertebrate species have decreased by an average of nearly 70% since 1970, and some species have vanished entirely or stand on the brink of extinction (“Home,” n.d.). This is a start, a reminder that the world we share with countless other species is nearing a critical tipping point

Over the past several decades, conservation has evolved into a global profession, uniting scientists, field biologists, and local communities in the effort to protect biodiversity. Estimates suggest that more than 200,000 professionals dedicate their work to understanding, monitoring, and safeguarding global ecosystems (Appleton et al. 2022). Despite the diversity of their disciplines, spanning forest ecology, ornithology, and marine biology, they share a common requirement: the need for data. Conservation moves at the speed of data. Reliable, timely, and accurate information is essential for tracking species populations, assessing policy impacts, identifying emerging threats, and evaluating the effectiveness of interventions.

Technological tools such as camera traps and passive bioacoustics recorders have transformed biodiversity monitoring by enabling continuous, non-invasive observation (Buxton et al. 2018). However, their deployment remains constrained by several persistent challenges: limited power supply, difficulty in data retrieval, equipment reliability, and restricted site accessibility (Pringle et al. 2025). Most devices rely on batteries and local storage, requiring researchers to periodically revisit deployment sites to collect data or replace equipment. These logistical burdens often confine monitoring efforts to areas accessible by road or trail, precisely the subset of ecosystems that are easiest to reach, rather than those most in need of sustained observation. Paradoxically, many of the world's most biodiverse and fragile habitats are also the most remote and inaccessible. As such, research in these areas often requires the construction of new infrastructure to reach them, which can further undermine efforts to protect these ecosystems (Kleinschroth et al. 2017).

Even when data can be successfully collected, much of what the sensors capture is not useful. The noise-to-information ratio can be extremely high, requiring conservationists to spend significant time reviewing and filtering out irrelevant data. Camera traps help to partially mitigate this challenge by using infrared (IR) sensors to trigger image capture only when movement is detected, substantially improving the signal-to-noise ratio of collected data. However, they still capture a significant number of blank images due to false triggers. In contrast, bioacoustics monitoring and other continuous recording methods typically lack an equivalent event-triggering mechanism. This absence of a selective trigger creates a costly and time-consuming workflow that limits both the scalability and timeliness of biodiversity monitoring, particularly in remote, data-poor ecosystems where sustained observation is most urgently needed (Kershenbaum et al. 2025). Overcoming these challenges requires a new generation of monitoring tools that are autonomous, sustainably powered, and capable of real-time data processing and transmission without the need for constant human intervention.

Recent advances have introduced solar-powered systems, Low-Earth-Orbit (LEO) satellite, and global systems for mobile communications (GSM) connected camera traps, extending the reach of remote ecological monitoring. With these advancements, we present SPARROW, an open-source, modular system, comprising both hardware and software, for autonomous biodiversity monitoring. It can interface with a wide range of devices, including camera traps, bioacoustic recorders, and environmental sensors. The system employs a low-voltage edge GPU to process data directly from these sensors, enabling the execution of state-of-the-art AI classification models for on-edge analysis, and transmits results via LEO satellite or GSM communication. Designed to be affordable, reproducible, and robust, SPARROW supports continuous environmental monitoring in remote or infrastructure-limited regions.

The objectives of this paper are threefold:

1) To describe the open-source hardware and software architecture of the SPARROW system.
2) To evaluate its field performance under remote deployment conditions; and
3) To promote community-driven reproducibility and innovation through the public release of all materials, schematics, and code.

By bridging advances in AI, edge computing, and sustainable hardware design, SPARROW seeks to democratize access to next-generation biodiversity monitoring tools and to accelerate conservation science on a global scale.

## 2. Background

Biodiversity monitoring has advanced significantly over the past two decades using automated sensing technologies such as camera traps and bioacoustics recorders. Advances in mobile connectivity (3G/4G) and the integration of solar-powered systems have further enabled long-term monitoring in remote locations. However, significant challenges remain, including the cost of individual SIM cards, the inability to monitor extremely remote locations, and the continued need for manual data retrieval and analysis. The number of camera traps deployed in a study varies, with an average deployment of 78 cameras (Yong et al., 2018). A similar pattern is observed in bioacoustics monitoring, where large numbers of recording devices are often deployed across extensive areas. As such, some studies generate millions of images and recordings (McShea et al. 2016), creating substantial bottlenecks in data collection and processing that can delay or even prevent project completion.

The development of AI tools for conservation applications has received significant attention from the research community in recent years. Much of this work focuses on training accurate computer vision models capable of detecting and classifying diverse species in complex and high variance environments. Models and tools developed by researchers such as (Ahumanda et al. 2020), TrapTagger WildEye and PyTorch Wildlife (Hernandez et al. 2024) have transformed camera trap analysis by semi-automating the process through the application of deep learning. Similarly, bioacoustics platforms such as ARBIMON ("Arbimon – Empower Your Wildlife Research," n.d.), Rainforest Connection (Rainforest Connection, n.d.), and AudioMoth (Open Acoustic Devices, n.d.-a) have expanded the reach of acoustic biodiversity monitoring, enabling the collection of long-term datasets across diverse ecosystems. Complementary audio processing tools and machine learning frameworks, including BirdNET (Kahl et al., 2021) and Arbimon's automated analysis pipelines, have further advanced the field by facilitating large-scale species detection and acoustic pattern recognition.

Despite their success, these systems often rely on battery power, limited communication networks and manual data retrieval, limiting their scalability and deployment in remote or infrastructure-poor environments. Moreover, the absence of on-device (edge) processing requires that all data be either manually retrieved or transmitted over costly and unreliable networks, further limiting their operational efficiency.

As such, researchers have increasingly sought to integrate AI with edge computing to reduce data processing burdens and improve the timeliness of biodiversity monitoring. Recent work has demonstrated the feasibility of deploying deep learning models directly on low-power embedded hardware for species detection and behavioral monitoring through systems such as WildDrone (WildDrone – Drones for Wildlife Conservation, n.d.), and Conservation X Labs ("Conservation X Labs," n.d.). These advances highlight the potential of decentralized, automated analysis but remain constrained by limited power autonomy and network connectivity, particularly in regions lacking GSM coverage. Parallel developments in environmental sensor networks and the Internet of Things (IoT) have produced solar-powered, remotely connected systems for ecological and environmental monitoring. Projects such as FieldKit (FieldKit, n.d.) and Open Acoustic Devices (Openacousticdevices, n.d.-b) have championed open-source, modular architectures that foster community-driven innovation and reproducibility. However, few existing systems combine solar power, on-edge AI computation, and satellite connectivity within a unified, open platform specifically designed for biodiversity monitoring.

SPARROW builds upon these developments by integrating solar-powered energy systems, low-voltage GPU computing, and flexible communication options, including both GSM and LEO satellite uplinks—into a single open-source framework. By uniting visual, acoustic, and environmental sensing within one modular architecture, SPARROW addresses long-standing barriers to sustained data collection in remote ecosystems. Its open design aims to democratize access to advanced biodiversity monitoring technologies and promote reproducibility, scalability, and real-time environmental intelligence in regions that have historically been data-poor or logistically inaccessible.

## 3. System overview

SPARROW functions as an autonomous, self-sustaining monitoring node engineered for deployment in remote and environmentally challenging locations. The system is housed within a waterproof and weather-resistant enclosure that protects its core components. A SPARROW unit is comprised of the following core components:

- On-edge computation: A Raspberry Pi 5 (4-core Arm Cortex-A76 CPU at 2.4 GHz with 8GB of RAM) or an NVIDIA Jetson Orin Nano (6-core ARM Cortex-A78AE CPU and a 1024 CUDA core NVIDIA Ampere GPU) performs on-device processing of visual, acoustic, and environmental data through embedded deep learning models.
- Local wireless transmitter (e.g., Wi-Fi, Xbee, and LoRa): Functions as an access point to form a local mesh network, enabling nearby sensors and devices to transmit data directly to the SPARROW node.
- Power system: Solar panels, charge controller and battery provide continuous off-grid power. The charge controller and battery are instrumented for real-time monitoring and control, enabling adaptive power management and system reliability.
- Global connectivity: A hybrid communication system that supports both LEO satellites and GSM to provide reliable remote data transmission.

SPARROW integrates multiple environmental sensors, including temperature, humidity and air-pressure monitors, and supports connections to peripheral devices such as camera traps, video recorders, and bioacoustic microphones. Data from these devices are managed by the SPARROW Client, a modular software framework implemented as a set of containerized microservices deployed on the Raspberry Pi 5 or NVIDIA Jetson Nano Orin. Using PyTorch Wildlife (Hernandez et al. 2024), an open-source library for on-device inference, SPARROW performs real-time data processing and local AI-based detection and classification directly in the field. Optimized deep learning models are deployed through NVIDIA Triton Inference Server, enabling efficient, low-latency inference. This configuration allows SPARROW to identify species, detect behavioral and environmental patterns, or trigger automated events without requiring constant connectivity to external servers.

Once data is processed, summarized information and relevant media files are uploaded to the *SPARROW server*, a centralized platform that aggregates inputs from multiple field units for further analysis and visualization. To optimize energy efficiency, SPARROW employs a dynamic communication schedule: it can batch and transmit data periodically (e.g., once per day) to conserve power, while maintaining the ability to “wake” the uplink module to send immediate alerts when predefined classification rules or anomaly thresholds are met. Figure 1 (a) shows the basic system architecture of the SPARROW device.

This integrated architecture enables SPARROW to operate as a robust, self-sustaining biodiversity monitoring system capable of autonomous data acquisition, processing, and transmission in regions with limited power and network infrastructure. Figure 1 (b) shows an example of deployment of a SPARROW device.

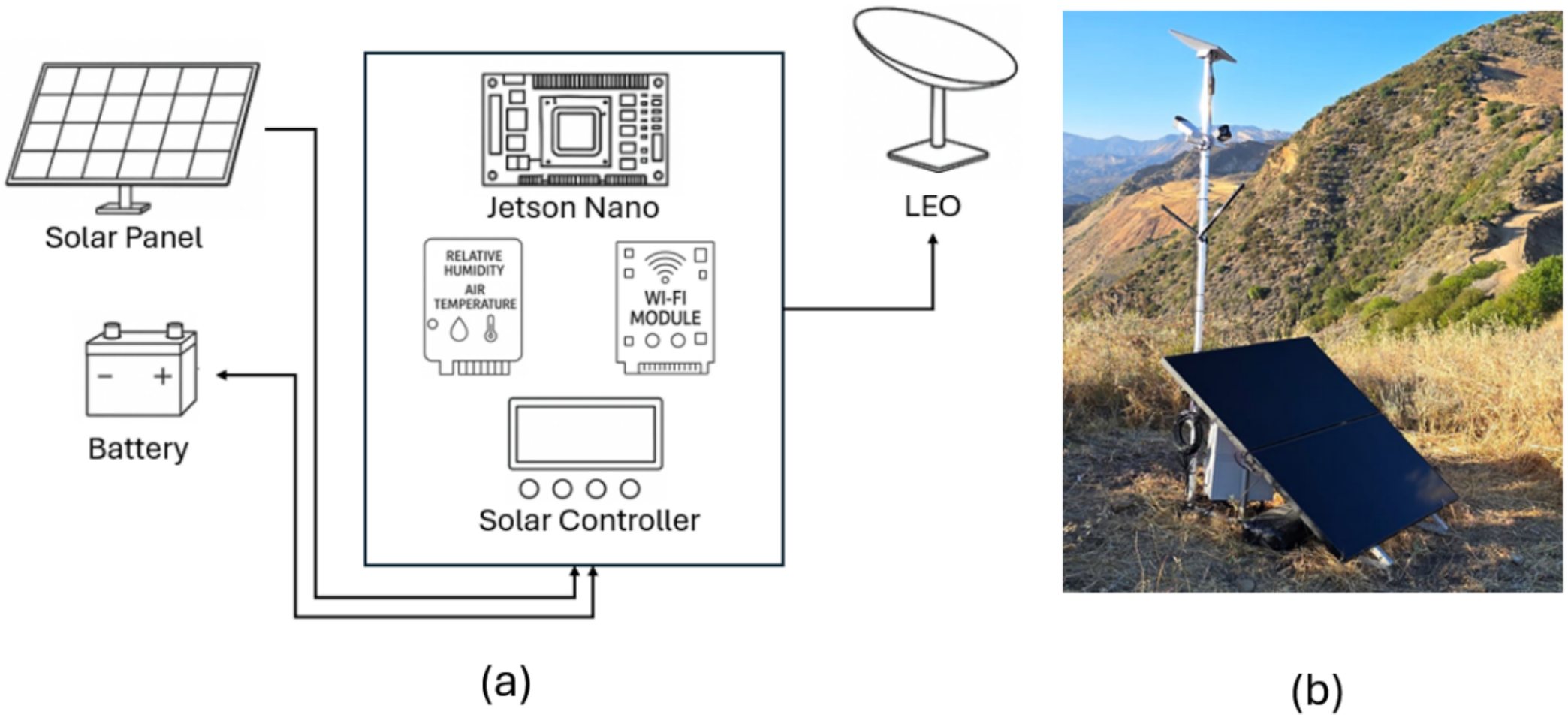

*Figure 1. (a) System architecture of the SPARROW system. (b) Sparrow field deployment.*

# 4. Hardware design

The SPARROW hardware platform was designed to operate autonomously in remote infrastructure limited and environmentally challenging locations. Its architecture emphasizes modularity, energy efficiency, and resilience to extreme weather conditions, while maintaining a low-cost, reproducible design that can be assembled from commercially available components. These principles align with established best practices for modular IoT platforms in environmental monitoring, which prioritize adaptability, fault tolerance, and sustainability in harsh conditions (Liao et al., 2025; Vasconcelos & Nunes, 2022), allowing the system to operate across a wide range of ecological conditions while remaining reproducible and maintainable in the field.

Rather than treating sensing, compute, and communication as isolated subsystems, SPARROW integrates them into a unified architecture where data acquisition, processing, and transmission are integrated and tightly coupled. A centralized architecture enables pushing substantial portions of the processing pipeline "to the edge" (i.e., co-located with sensors), rather than relying on a strictly "collect-retrieve-process" workflow. This architectural model coupled with open-source hardware and energy-efficient platform designed for scalable deployments in conservation contexts, underscores the importance of the role of edge computing in reducing the latency and power consumption from biodiversity monitoring efforts (Kline, n.d.).

SPARROW is implemented as a small family of validated hardware configurations, rather than a single fixed bill of materials. In this manuscript, we define the Raspberry Pi 5-based system as the reference build, because it is the lowest cost and easiest configuration to procure globally, assemble, provision, and maintain in conservation field contexts. This decision reflects a pragmatic deployment constraint repeatedly encountered in real-world collaborations: although embedded GPU platforms provide additional compute headroom (e.g., more computationally intensive real-time video processing), the operational friction, higher cost, and availability constraints of maintaining GPU-centric stacks can slow adoption by partner teams. This rationale is informed by operational experience and lessons learned from field deployments across geographically and culturally diverse environments.

The Jetson Orin Nano configuration is therefore treated as a secondary, performance-oriented variant for deployments where on-device workloads exceed what is practical on the reference Raspberry Pi configuration (e.g., more demanding real-time pipelines).

## 4.1 Core Unit

At the heart of SPARROW is a Respberry Pi 5 single board computer (SBC), which provides low-power, high-performance computing capabilities for on-edge AI inference. The Raspberry Pi 5 enables real-time processing of images, audio, and environmental sensor data using deep learning models implemented in PyTorch or other deep learning frameworks. The processing unit is housed within a weatherproof IP65/Nema 4(x) rated, ensuring durability in tropical, desert, and forest outdoor environments.

The rationale for selecting a Pi-class SBC as the reference is motivated by the fact that in biodiversity monitoring, edge AI deployments are inherently constrained by coupled tradeoffs between power, compute, and communication. Practical deployments require integrated co-design spanning sensing modality, embedded compute, model optimization, and wireless networking. As a result, "reference" hardware must reflect what deployment teams can realistically source, provision, and operate at scale, not merely what maximizes peak inference throughput (Vuilliomenet et al., 2026). In camera-trap edge inference specifically, prior work evaluating embedded devices shows that Raspberry Pi–class SBCs can support end-to-end edge classification architectures, while underscoring the importance of optimizing models and pipelines for resource-limited operation (Zualkernan et al., 2022)

### 4.1.1 Performance variant (optional GPU enabled module)

While Raspberry Pi 5 is the reference platform, SPARROW also supports a GPU-accelerated compute variant for deployments requiring higher sustained inference throughput or more computationally intensive models. One supported option is the NVIDIA Jetson Orin Nano series, configured to a 7 W power profile. This compute class can be appropriate when the workload cannot be met by the reference SBC while maintaining required duty cycles.

Empirically, camera-trap edge literature shows that Jetson-class devices can deliver substantially lower per-image latency than a Raspberry Pi in stress-testing scenarios, but typically at higher power draw, reinforcing the rationale for treating GPU compute as a variant rather than the default baseline (Zualkernan et al., 2022).

## 4.2 Connectivity and Communication

SPARROW supports multiple communication modes to operate effectively across diverse connectivity conditions. Local device connectivity (e.g., cameras, acoustic recorders, and sensors) is supported via Wi-Fi and wired interfaces, enabling short-range data transfer to the hub. For remote deployments, SPARROW can be configured with long-range communication, including cellular (GSM/LTE, where available) and satellite uplinks services such as low-Earth-orbit satellite connectivity (e.g, Starlink). This multimodal design supports operation across settings ranging from infrastructure-rich field stations to remote rainforest and savannah deployments, where connectivity is intermittent or absent.

When a low-Earth-orbit satellite terminal is used, power budgeting is accounted for and managed directly by the main compute system.

## 4.3 Sensor Integration

SPARROW integrates a comprehensive ultra-low power sensor array for high precision environmental monitoring and contextual metadata, including temperature, atmospheric pressure, and humidity sensors, along with advanced gas sensing capabilities for air quality analysis. It also interfaces with external modules such as camera traps, acoustic recorders, and motion sensors through standard USB or GPIO connections. The modular architecture is intentionally designed to accommodate heterogeneous sensor combinations without redesigning the core unit, enabling deployment-specific tailoring across ecological applications, from acoustic biodiversity monitoring to environmental hazard detection. Figure 2a. Shows the Raspberry Pi 5 with the Sensor expansion board. Figure 2b. Shows the Jetson Orin Nano with the Sensor expansion board.

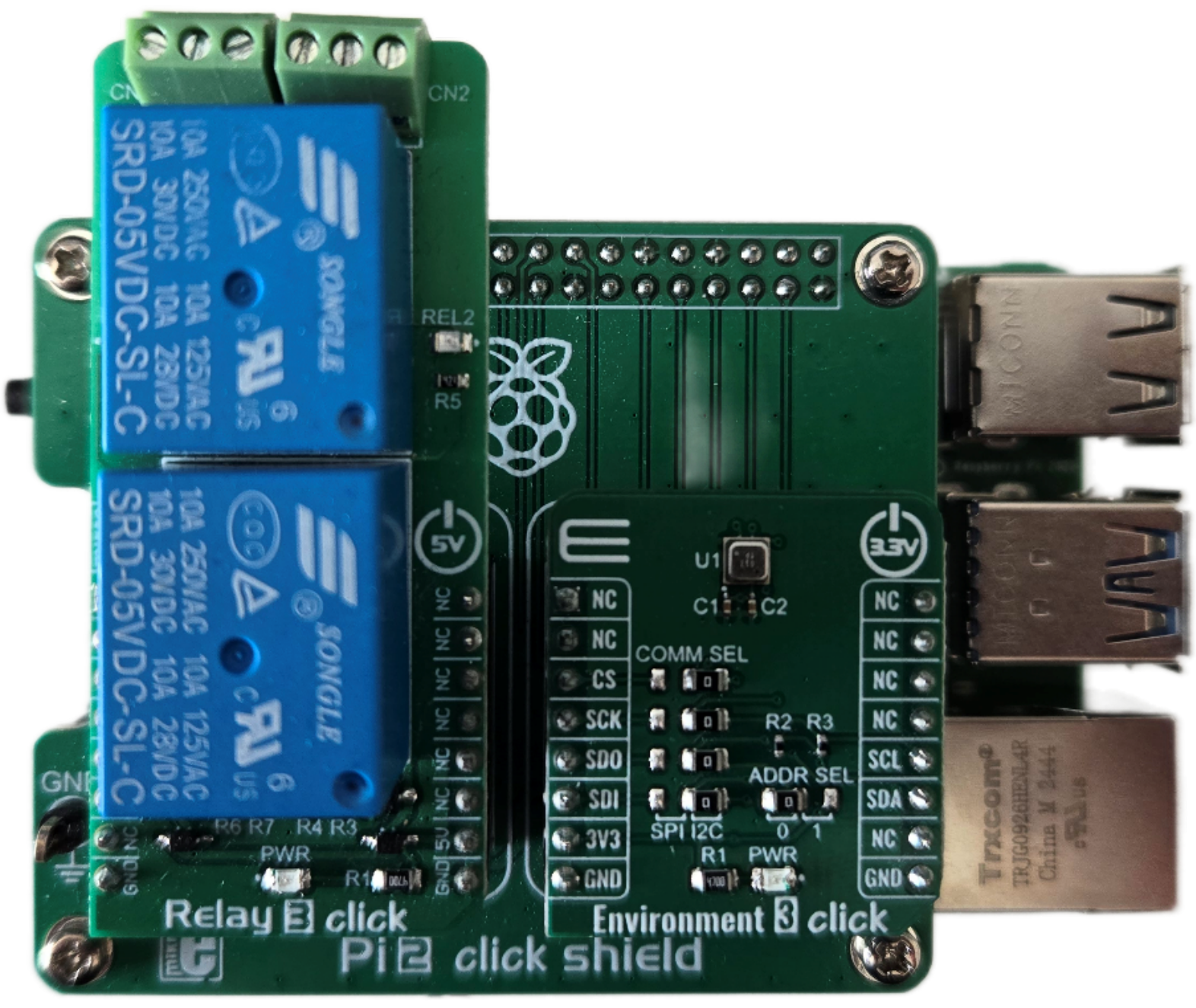

*Figure 2a. Raspberry Pi based SPARROW Compute unit with Core sensor expansion board.*

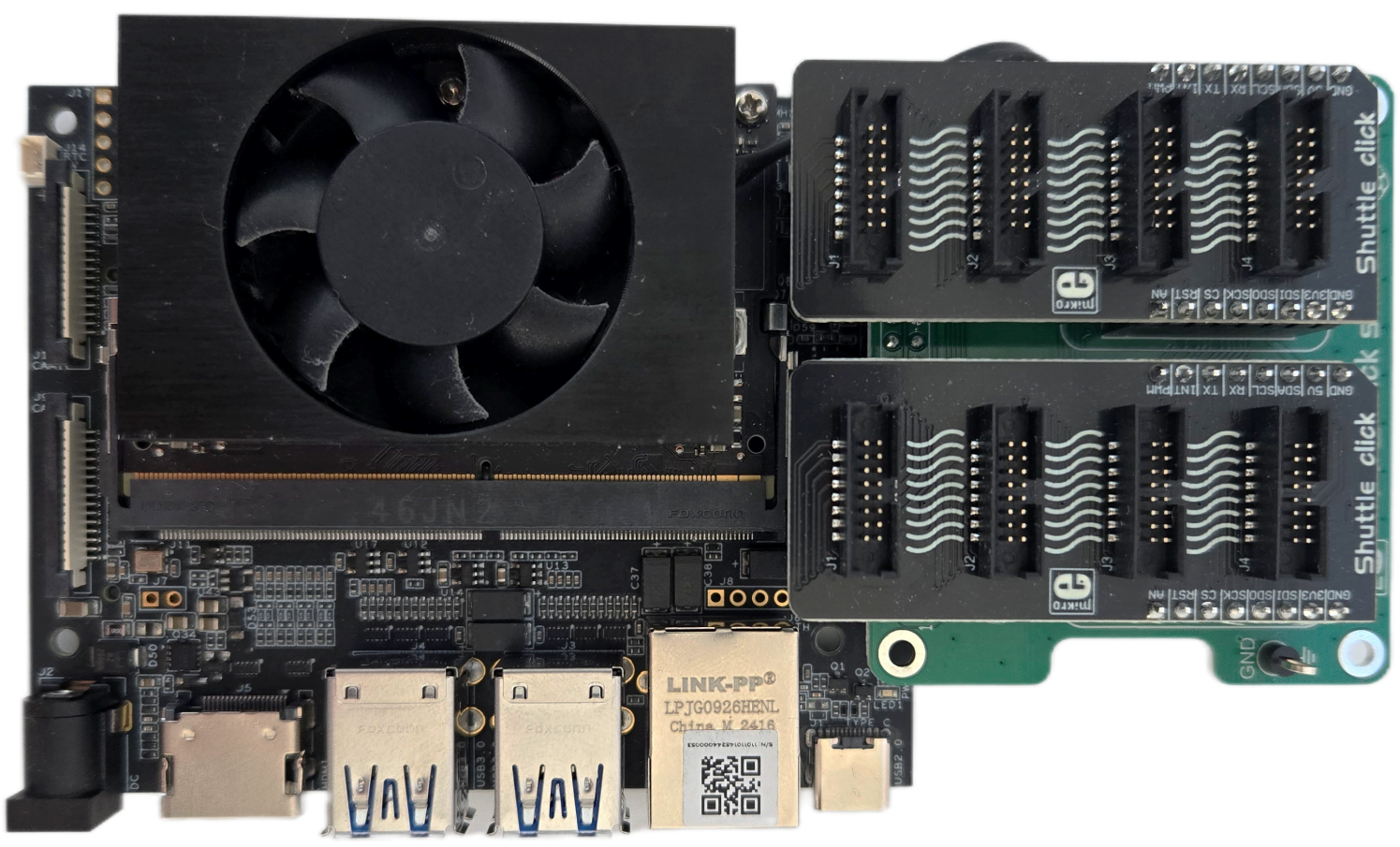

*Figure 2 b. Jetson Orin Nano based SPARROW Compute unit with Core sensor expansion board.*

## 4.4 Modular Design and MikroBUS Integration

A modular design approach was chosen to maximize flexibility and scalability while minimizing complexity. SPARROW incorporates the MikroBUS interface as a standardized expansion mechanism, enabling plug-and-play attachment of sensors and peripheral modules and supporting rapid reconfiguration between deployments. This standard lowers the barrier for users without deep hardware expertise by enabling plug-and-play compatibility and straightforward maintenance. Its practicality is well-documented in academic and engineering contexts, including educational demo systems (Multi-functional Demo Module, 2018), rapid prototyping of low-power sensor nodes (Design and Implementation..., 2023), payload development for space applications (Innovative Payload Interface Board, 2023), and modular sensor platforms (ModuLab, 2025).

The MikroBUS standard was selected because it offers a uniform interface that simplifies sensor integration, allowing easy swapping and expansion without redesigning the core system.

## 4.5 Enclosure and Field Deployment

The physical layout of SPARROW prioritizes ease of assembly, durability, and long-term maintenance in harsh environments. Core electronics (compute, power regulation, and sensor subsystems), are mounted and housed within an IP65-rated sealed enclosure (as per IEC IP code framework IEC 60529) to protect against dust, insect and water ingress. All enclosure penetrations (power, antennas, sensors) are implemented via sealed cable glands to preserve the ingress boundary while supporting deployment-specific wiring and replacement in the field.

Externally exposed subsystems such as the $LiFePO_4$ battery pack, solar panels, satellite antenna, and Wi-Fi antennas are mounted outside the sealed enclosure
so that high-wear components can be replaced without opening the electronics enclosure.
Thermal management prioritizes passive strategies including mesh-filtered cooling vents to prevent debris and insect accumulation while ensuring system thermal stability.

All structural elements use corrosion-resistant materials to withstand prolonged exposure to humidity, and UV radiation. SPARROW's modular layout enables quick field servicing and component replacement without the need for specialized tools, making it ideal for remote infrastructure-poor environments where resilience and autonomy are critical.

## 4.6 Power System and duty-cycled operation

SPARROW is designed for off-grid autonomy using a solar-battery architecture sized around continuous sensing/compute plus duty-cycled communications. The standard energy system is configured using a pair of 100W 12V solar panels working in a 24V configuration (in series) and a 24 V, 100 Ah lithium-iron-phosphate ($LiFePO_4$) battery, providing roughly 2.4 kWh of nominal storage capacity. The reference Raspberry Pi 5 SBC draws approximately 2.7–3.0 W at idle and can reach ~12 W peak under intensive workloads (Schweizer, 2026; Flaherty, 2023). For image inferencing on Raspberry Pi 5–class edge hardware, a ~5 W average is a conservative budgeting assumption, consistent with published camera-trap edge deployments reporting ~0.84 A average current on Raspberry Pi during sustained processing (≈4.2 W at 5 V) (Zualkernan et al., 2022).

The NVIDIA Jetson Nano typically draws 5–10 W depending on workload (NVIDIA, n.d.), and the solar controller with auxiliary sensors adds approximately ~2 W. The Starlink Mini uplink, used primarily for data transmission, consumes 20–25 W but is only powered for a limited duration, typically one hour per day, to conserve energy. Under typical conditions, the total daily energy demand (~300 Wh) remains well below the generation capacity of the solar system (800–1000 Wh/day), providing a strong operational margin and enabling continuous 24×7 sensing and analysis.

*Table 1. SPARROW power requirement breakdown*

| Component | Average Power Use (Wh) | Duty Cycle | Daily Energy Use (Wh) | Notes |
|---|---|---|---|---|
| Raspberry Pi 5 | 5 | 24 h/day | 120 | Includes on-edge inference and idle periods |
| Jetson Nano | 7.5 | 24 h/day | 180 | Includes on-edge inference and idle periods |
| Sensors | 1.0 | 24 h/day | 24 | Temperature, pressure, and other sensor modules |
| Starlink Mini | 22.5 | 1 h/day | 22.5 | Scheduled daily data upload |
| Wi-Fi Local Network | 2.0 | 24 h/day | 48 | Local device communication |
| **Total (approx.)** | — | — | **~200-300 Wh/day** | — |
| **Solar Generation Capacity** | **200 W panel** | ~5 h effective sunlight/day | **~800–1000 Wh/day** | Surplus supports cloudy periods |

## 4.7 Power Management

SPARROW employs an adaptive power management strategy orchestrated by the solar charge controller and onboard firmware. Core subsystems including the main compute, sensor array, and Wi-Fi, operate continuously to allow uninterrupted environmental monitoring and real-time edge inference. In contrast, the Starlink Mini uplink is conditionally activated based on a preconfigured schedule (e.g., a one-hour daily window) for data uploads and system synchronization. To safeguard energy reserves, the system automatically suppresses Starlink Mini activation when battery voltage falls below a predefined threshold, prioritizing essential sensing and local computational tasks.

Beyond the scheduled communication window, the system supports event-triggered wakeups: if a high-priority classification (e.g., detection of a specific species or anomaly) occurs, SPARROW can override the schedule and immediately initiate a Starlink Mini connection to transmit an alert. This combination of continuous monitoring, scheduled data transfer, and conditional uplink activation ensures robust autonomy, even under variable solar and challenging weather conditions.

## 4.8 Environmental Operating Range

SPARROW is engineered to operate reliably across a wide range of environmental conditions, from tropical rainforests to high-altitude or sub-arctic regions. All core components are housed within a sealed, IP65-rated waterproof enclosure, providing protection against rain, dust, and humidity. The system's thermal limits are primarily defined by its electronic and energy storage components. The main compute unit and solar charge controller can operate between –25 °C and +80 °C, while the $LiFePO_4$ battery functions optimally between –10 °C and +60 °C, with charging efficiency decreasing below freezing.

For cold-weather deployments, SPARROW employs a thermal co-location design, grouping the $LiFePO_4$ battery, main compute unit, and other heat-emitting electronics within an insulated compartment. Residual heat from the main compute workloads help maintain an internal temperature above 0 °C, preventing battery degradation and ensuring stable performance during cold nights. This passive heat retention, combined with internal temperature monitoring and automatic throttling, allows SPARROW to remain fully operational even when ambient temperatures drop well below freezing.

In field tests, SPARROW has demonstrated continuous operation in environments ranging from –10 °C to +50 °C, maintaining data integrity and uninterrupted on-edge processing. Future versions will extend this capability further through optional phase-change thermal buffers and low-power resistive heating elements for extreme cold-weather applications.

## 4.9 Hardware Cost

Although alternative components exist for most of the items in the Bill of Materials, the Jetson Orin Nano and Starlink Mini Kit do not have readily available substitutes that can be easily integrated into the system. The total estimated cost of a SPARROW main unit is ~2200 USD.

*Table 2. SPARROW main component cost breakdown.*

| Component | Unit cost (USD)* | Replaceable with alternative components? |
|---|---|---|
| Raspberry Pi 5 or Jetson Orin Nano | 175 - 249 | No |
| Brain Assembly (Storage, Sensors, Relay, etc.) | 306 | Yes |
| Power Assembly (Solar Power Controller, Battery, USB interface, etc.) | 999 | Yes |
| Starlink Mini Kit | 299 | No |
| Network Assembly (Ethernet Cable, Antennas, etc.) | 130 | Yes |
| Build Materials (Electrical Junction Box, Cables, Glands, etc.) | ~ 200 | Yes |
| **Total Cost** | ~ 2000 - 2200 | |

*The costs shown in the previous chart are expressed in U.S. dollars and represent component prices as sourced within the United States at the date of publication. These figures do not include shipping fees or applicable taxes.

## 5. Software Architecture

The SPARROW software architecture is organized into two primary components: the SPARROW Client, which operates on the edge device, and SPARROW Studio, the main user interface, which manages the centralized data aggregation, visualization, and configuration. Both are built around containerized microservices, providing modularity, scalability, and redundancy. This design allows individual services to be updated, replaced, or scaled independently without affecting system stability or requiring full redeployment.

The SPARROW Client runs on either a Raspberry Pi 5 or an NVIDIA Jetson Orin Nano embedded computing platform. The software stack is adapted to the underlying hardware architecture to balance performance and resource efficiency. On Jetson-based deployments, the system is configured with Ubuntu 20.04 LTS and the NVIDIA JetPack 5.x SDK, which provides integrated CUDA, cuDNN, and TensorRT libraries for GPU-accelerated edge inference. In this configuration, deep learning models are served locally using NVIDIA Triton Inference Server. On Raspberry Pi deployments, where GPU acceleration is more limited, inference is performed directly using ONNX Runtime for lightweight CPU-based execution. In both configurations, the client orchestrates local computation, device control, sensor integration, and communications. Figure 3 presents the overall software architecture, showing both the client and server software components.

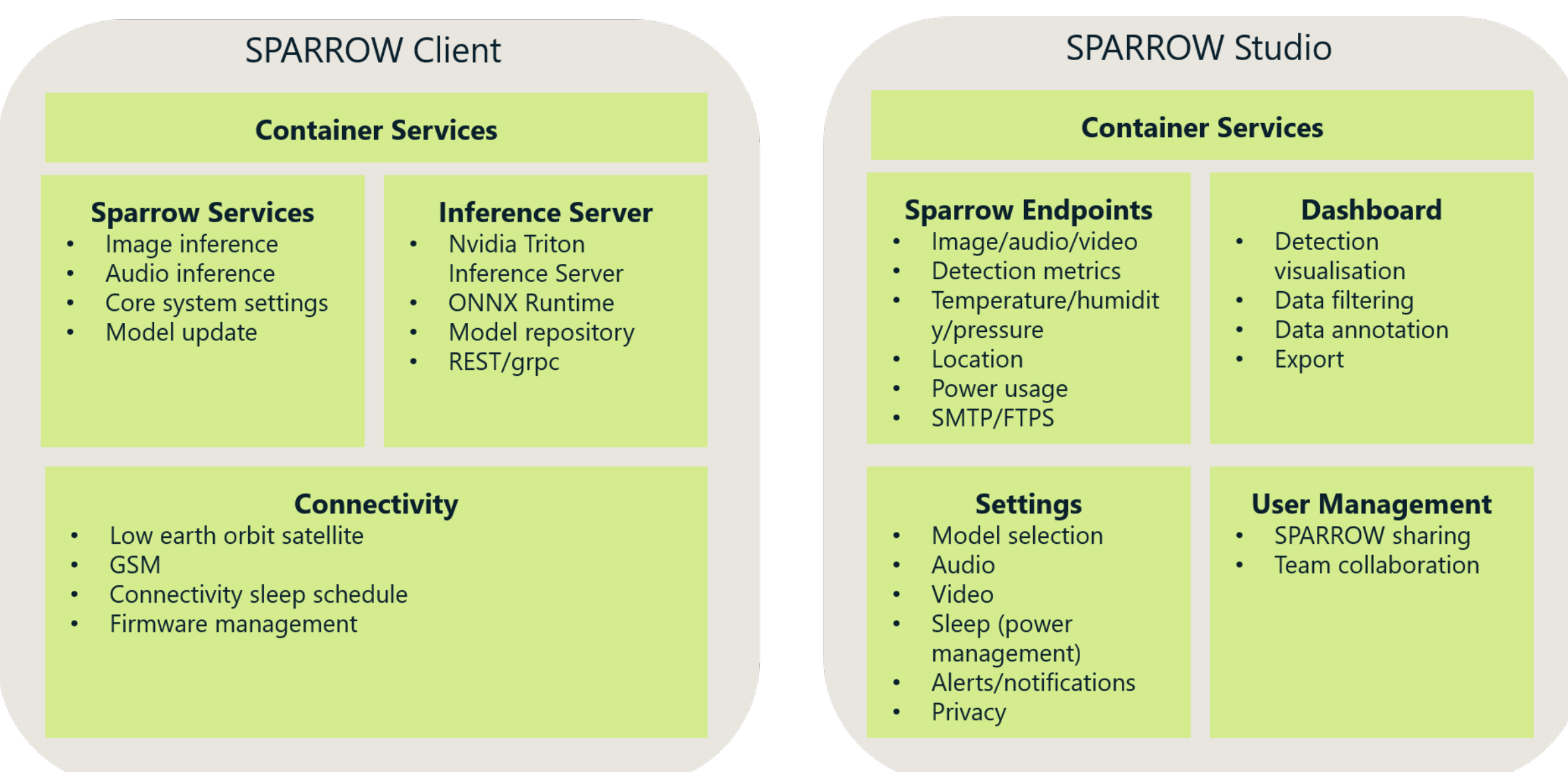

*Figure 3. Software architecture of the SPARROW platform.*

The SPARROW services container forms the central orchestration layer responsible for on-device data processing and analysis. The platform performs image and audio inference using locally hosted deep learning models, with the inference backend varying depending on the deployment hardware. Image data are processed using the MegaDetector V6 object detection model, followed by species-level classification using lightweight convolutional neural networks such as the AI4G Amazon classifier. The service supports configurable detection thresholds and optional filtering of blank images. A parallel audio processing pipeline performs spectrogram generation and classification, enabling real-time bioacoustics event detection. On Jetson platforms, these operations can leverage GPU acceleration through Triton and CUDA-enabled libraries, while Raspberry Pi deployments execute optimized ONNX Runtime inference pipelines. Model configurations and operational parameters are synchronized periodically from the SPARROW server through JSON configuration files.

The connectivity services container manages outbound communication across GSM, Wi-Fi, Ethernet, or LEO satellite networks, depending on the deployed hardware configuration and available infrastructure. The service implements device scheduling and bandwidth optimization strategies to support asynchronous data transmission, telemetry reporting, and remote management under intermittent or bandwidth-constrained connectivity conditions. It also supports over-the-air (OTA) synchronization of configuration files, firmware, and AI models, ensuring consistency across distributed deployments.

The inference subsystem is hardware-dependent and adapts to the computational capabilities of the deployment platform. On NVIDIA Jetson-based deployments, the platform uses NVIDIA Triton Inference Server with a versioned local model repository. Triton exposes standardized HTTP/REST and gRPC interfaces for inference requests originating from the SPARROW services container, enabling low-latency GPU-accelerated execution. A background synchronization process periodically polls the SPARROW server for updated models, downloads new model versions, and reconciles the local repository through Triton's repository management interface.

On Raspberry Pi Raspberry Pi deployments, where GPU acceleration and memory resources are more constrained, models are executed directly using ONNX Runtime without a dedicated inference server layer. In this configuration, models are loaded and managed locally by the SPARROW services container to minimize overhead while maintaining support for configurable and remotely synchronized AI pipelines.

The platform additionally supports user-defined model deployment, enabling researchers to load, host, and evaluate custom deep learning models directly on edge devices across both deployment architectures.

## 5.1 SPARROW Studio

SPARROW Studio serves as the central server and management platform for the SPARROW ecosystem, supporting data ingestion, device orchestration, AI-assisted analysis, and collaborative review workflows. The platform is designed for both traditional camera-trap surveys and near real-time ecological monitoring deployments using connected edge devices and wireless field cameras.

Data can be ingested from multiple sources, including GSM and 4G-enabled camera traps using SMTP or FTPS transmission, manual batch uploads from local archives, and SPARROW edge devices deployed in the field. Uploaded images, audio, and telemetry are automatically processed using AI pipelines for wildlife detection and optional species classification before being organized into project workspaces for review and analysis. Figure 4 shows the SPARROW Studio dashboard.

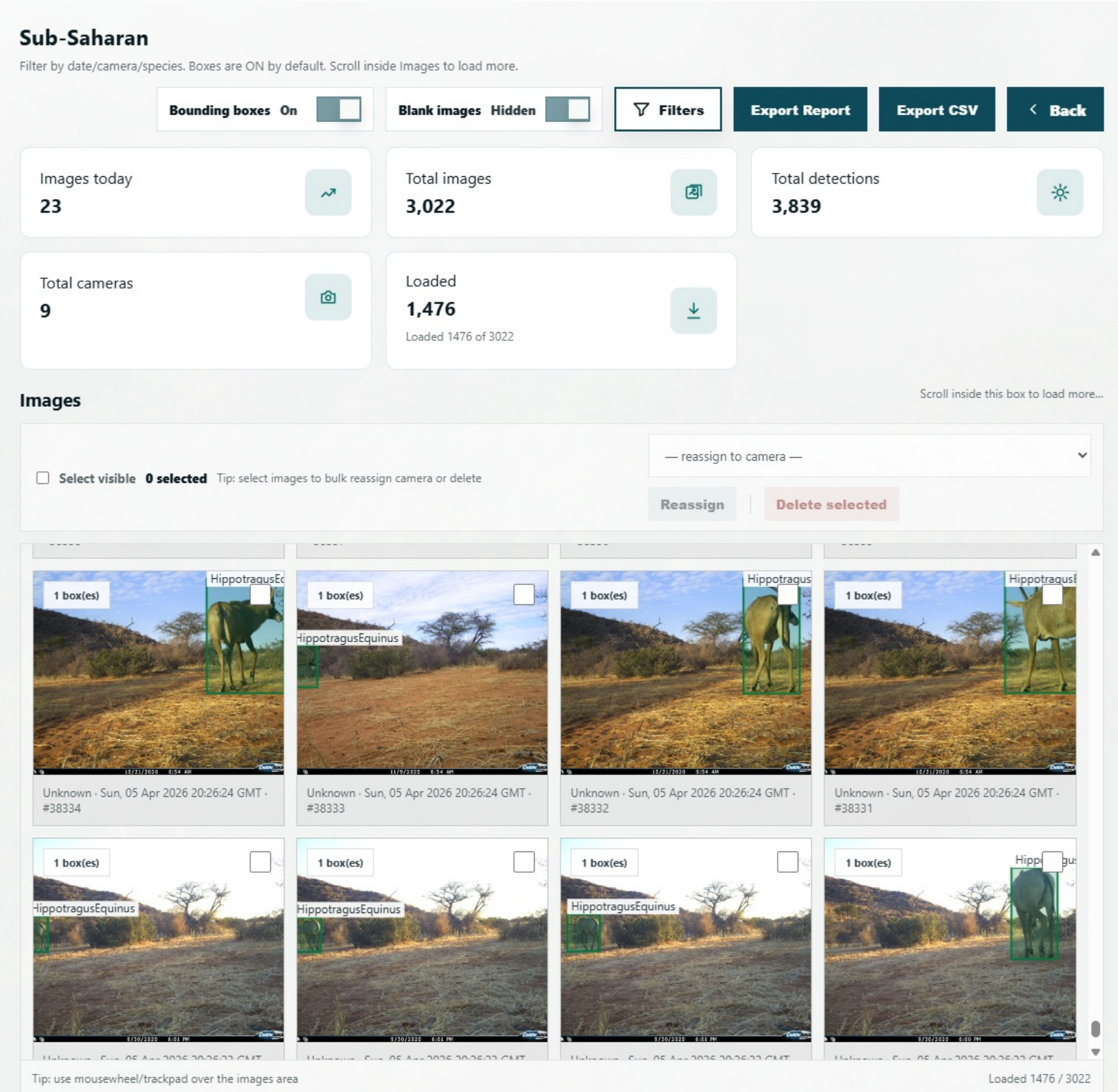

Figure 4. SPARROW dashboard overview showing projects and recent detections.SPARROW Studio provides a collaborative review environment where researchers can validate detections, correct species labels, remove false positives, and annotate observations. Images are displayed with detection bounding boxes and species labels overlaid directly onto each image, as shown in Figure

5. Filtering and batch-processing tools enable efficient navigation of large biodiversity datasets by species, deployment, camera, date range, or review status.

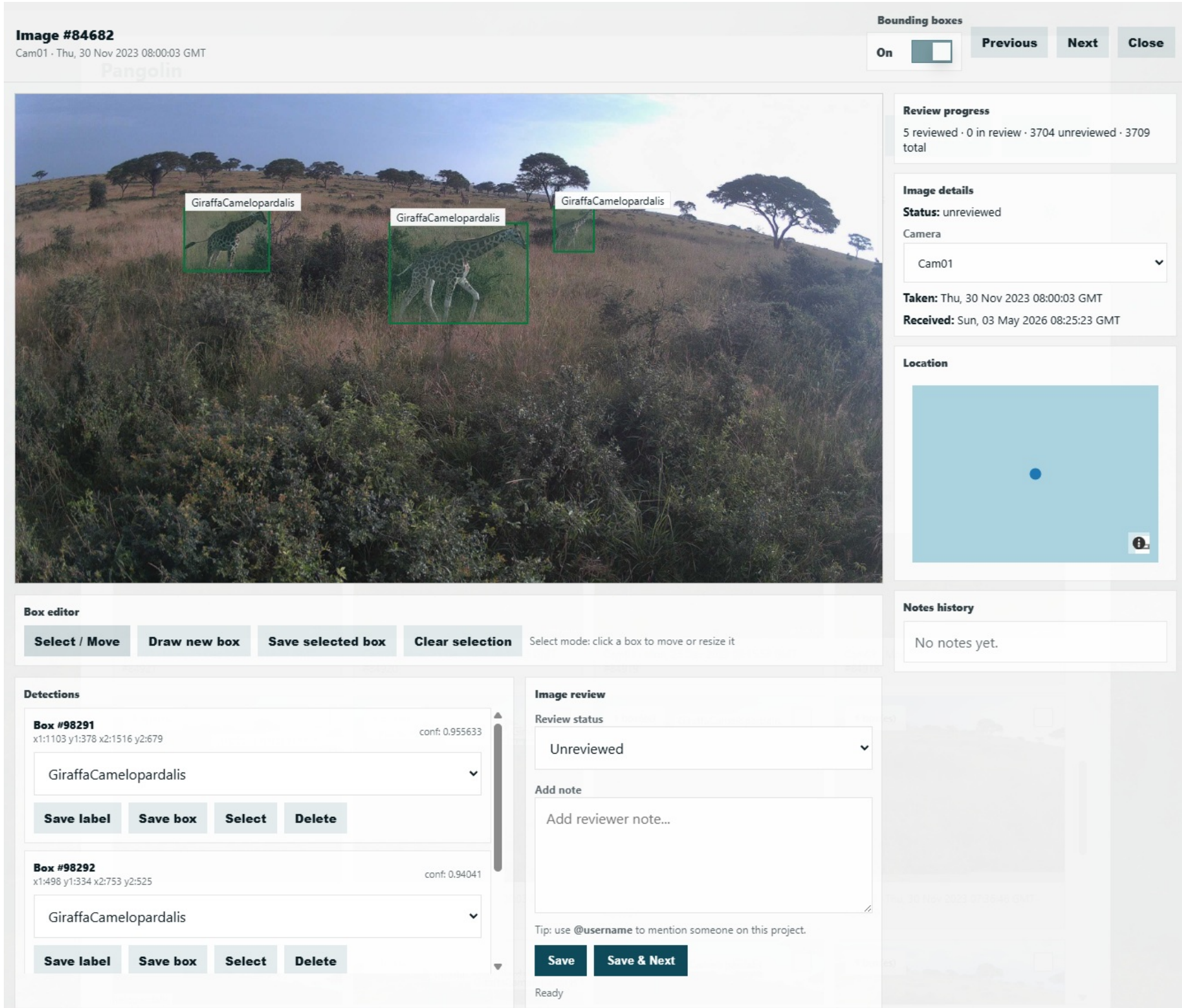

*Figure 5.* Review *workspace with detection thumbnails and overlays.*

The platform supports both real-time and offline survey workflows. Historical datasets can be uploaded through browser-based batch import tools, while connected GSM or 4G camera traps can automatically transmit images through monitored email inboxes or FTPS endpoints for near real-time monitoring applications. This enables rapid ecological assessment and automated alerting for target species detections.

SPARROW Studio additionally manages biodiversity metadata and deployment information required for long-term ecological studies and standards-compliant data export. Camera deployment records can store metadata including deployment dates, habitat information, camera orientation, mounting height or water depth, bait use, and deployment notes. The platform follows the Camtrap DP biodiversity standard to support interoperability with external biodiversity repositories and ecological data archives.

Beyond image review, the platform includes biodiversity reporting, notification systems, and collaborative project-management capabilities. Role-based permissions support administrators, project managers, reviewers, and viewers, enabling multiple research teams to collaborate within the same survey environment. Automated reporting tools generate biodiversity summaries containing species detections, independent detection counts, survey statistics, and per-camera activity metrics. Configurable species-specific alerts additionally allow users to receive notifications when target species are detected in near real time.

# 6. SPARROW Variants

## 6. 1 Edgeless SPARROW

When using commercially available wireless 4G-enabled camera traps, SPARROW has a lightweight edgeless deployment variant designed for quick deployment. Unlike the full SPARROW edge platform, which performs on-device AI inference, the lightweight SPARROW variant focuses on near real-time data acquisition from cellular devices (e.g., 4G cameras) and centralized cloud-based processing through SPARROW Studio.

The system supports camera traps capable of transmitting images and video over cellular networks using either SMTP email delivery or secure FTPS uploads. Images captured in the field are transmitted directly to SPARROW Studio, where they are automatically ingested, indexed, and processed using server-side AI pipelines for wildlife detection and optional species classification. This architecture allows existing wireless camera hardware to integrate into the SPARROW ecosystem without requiring dedicated edge-computing hardware at the deployment site. Figure 6 shows an example edgeless SPARROW deployment.

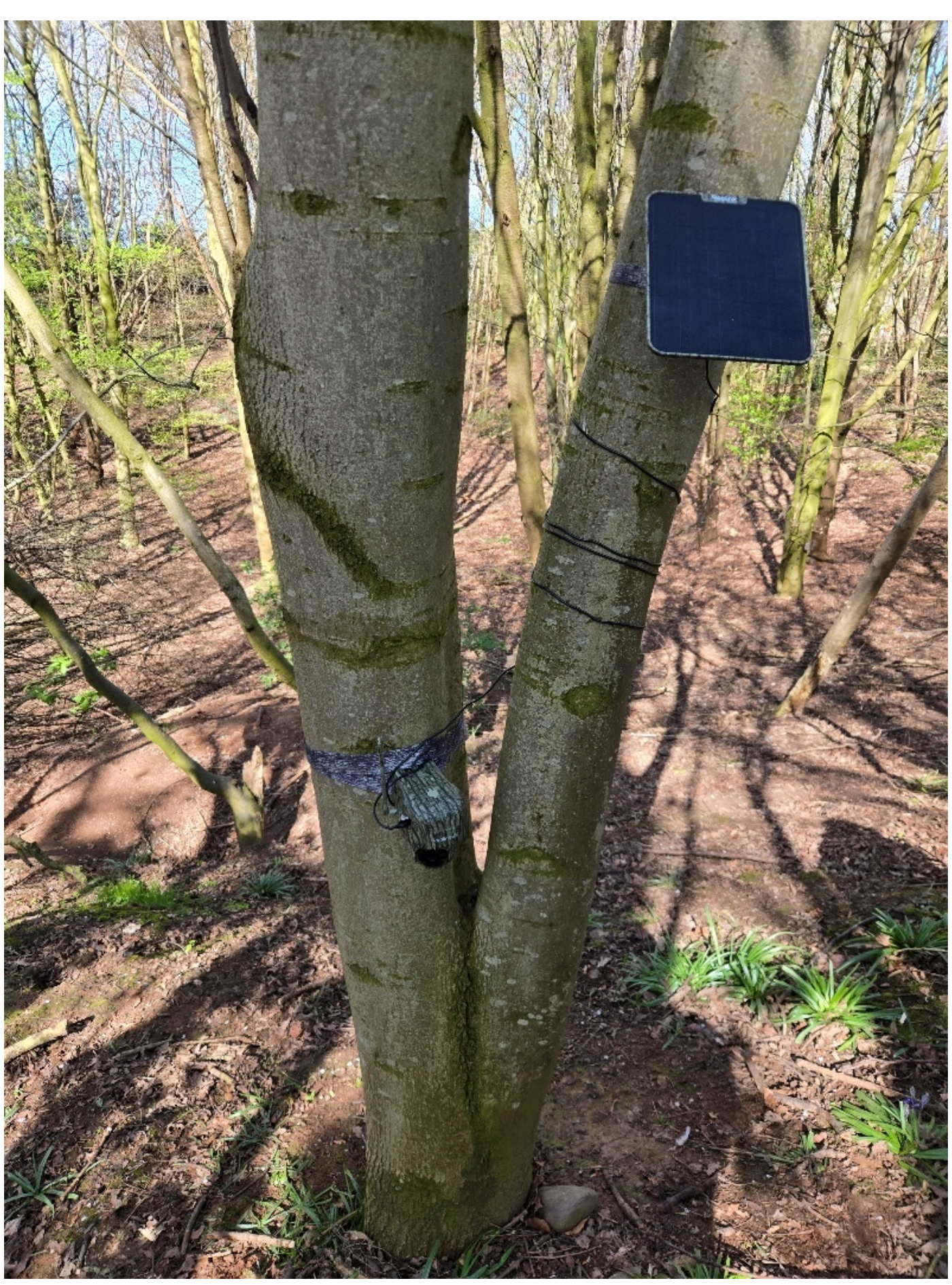

*Figure 6. Edgeless SPARROW deployment UK.*

Edgeless SPARROW is particularly suited for rapid-response monitoring deployments, temporary ecological surveys, remote wildlife observation, and situations where low-power field operation and simplified deployment logistics are prioritized over on-device inference. By leveraging existing 4G camera infrastructure, the platform enables near real-time biodiversity monitoring while maintaining access to the collaborative review, reporting, alerting, and data-management capabilities provided by SPARROW Studio.

### 6.2 SPARROW Mini

SPARROW Mini is a low-power variant of the SPARROW platform designed for long-duration and resource-constrained environmental monitoring deployments. In contrast to the full SPARROW edge system, which targets higher-performance edge AI workloads, SPARROW Mini prioritizes reduced power consumption, compact form factor, and simplified field deployment while still supporting near real-time biodiversity monitoring capabilities.

The platform is built around a Raspberry Pi Raspberry Pi Zero 2 W, paired with an integrated 4G communication module for remote telemetry and data transmission. This hardware configuration enables significantly lower energy consumption, making the system suitable for solar-powered, battery-powered, or off-grid deployments where long operational endurance is required.

SPARROW Mini can interface directly with existing camera traps, acoustic monitoring devices, or environmental sensors through USB connectivity or SD-card emulation workflows. Captured media are processed locally using lightweight AI inference pipelines based on the MegaDetector object detection model. Rather than performing computationally intensive species classification on-device, the system transmits detections, metadata, and selected media products to SPARROW Studio for centralized processing, review, and long-term storage.

To support distributed sensing deployments, each SPARROW Mini also integrates a XBee communication module, enabling the formation of low-power wireless mesh or point-to-point relay networks between multiple field nodes. In this architecture, secondary client nodes can forward detections, telemetry, and sensor data to a designated gateway node operating in server mode. The gateway node then aggregates and transmits the collected data to SPARROW Studio using the onboard 4G connection. This approach extends operational coverage in remote areas while reducing the number of required cellular uplinks and associated power consumption.

# 7. Field Deployment

## 7.1 SPARROW Deployments

Currently, fifteen SPARROW units have been deployed across seven distinct locations, reflecting a range of environments and collaborative partnerships. The full list of field deployments is presented in Table 3. As an example, the initial SPARROW deployment was at El Silencio Natural Reserve, located in Yondó, Antioquia, Colombia. Recognized for its highly diverse lowland rainforest habitats and role as a conservation corridor in the Magdalena River basin, the reserve was selected as a pilot site to demonstrate SPARROW's capabilities in long-term, multi-modal data collection. Six SPARROW units were installed during a two-week field campaign, each device powered by solar panels and connected via a satellite internet antenna strategically mounted in the forest canopy. Every SPARROW system was equipped with four cameras, two positioned in the canopy and two at ground level, to comprehensively capture wildlife activity across vertical strata. The primary monitoring objectives include supporting the reserve's management in wildlife assessment, tracking species of conservation concern, and informing restoration strategies. Future research will leverage this deployment to refine AI-based detection algorithms and expand collaborative ecological studies in tropical forest environments.

*Table 3. Current and planned SPARROW deployments worldwide.*

| Location | Biome | # of Units | Partners | Tasks | Deployment Date | Number of Images/Videos Collected | Hours of Audios collected |
|---|---|---|---|---|---|---|---|
| Colombia – Middle Magdalena | Tropical Rainforest and Wetland | 6 | Fundación Biodiversa Colombia | Animal Biodiversity Monitoring | 05/05/2025 | 5074 | 2191.5 |
| Seattle – Woodland Park Zoo | Temperate Forest | 1 | Woodland Park Zoo | Animal Monitoring | 05/21/2025 | 541 | 172.2 |
| California – Condor Reserve | Mediterranean Chaparral and Oak Woodland | 2 | National Geographic | Condor and Turtle Monitoring | 07/25/2025 | 90347 | 260.5 |
| California - Boucher Hill | Mediterranean Montane Forest | 2 | AlertCalifornia | Fire Alert | 07/25/2025 | 63992 | N/A |
| Colombia – Amazon Rainforest | Lowland Amazon Forerst | 2 | Instituto SINCHI | Animal Biodiversity Monitoring | 09/08/2025 | 438 | 105.9 |
| Perú - Amazon Rainforest | Lowland Amazon Forest | 1 | Los Amigos Biological Station | Predatory behavior Monitoring | 11/05/2025 | 1565 | 0.15 |
| Tanzania - Gombe National Park | Tropical Rainforest | 1 (6) | Jane Goodall Institute | Animal Monitoring | 24/08/2025 Part 2: Q1 2026 | N/A | N/A |
| DRC - Salonga National Park | Tropical Lowland Rainforest and Wetland | (12) | Salonga National Park and Virunga National Park | Bonobo Monitoring | Planned Q1 2026 | N/A | N/A |
| Seattle – Cascades National Park | Temperate Coniferous Forest | (1) | Woodland Park Zoo | Animal Monitoring | Planned Q4 2026 | N/A | N/A |

- Numbers in parentheses are planned number of units.

# 8. Discussion and Future Work

## 8.1 SPARROW Mini mesh network and compatibility to existing monitoring sites

The majority of existing SPARROW deployments have consisted of standalone main edge processing units or edgeless SPARROW configurations connected to 4G cameras. SPARROW has been designed with two key aspects that are now ready for deployment and testing:

1. Operation in truly remote areas for large-scale data collection and edge processing, using not only the main unit but also a SPARROW Mini mesh network.
2. Compatibility with existing non-wireless camera trap monitoring sites through direct integration with SPARROW Mini.

SPARROW systems incorporating the SPARROW Mini mesh network are currently being installed and tested in Uganda and the UK. These deployments will help finalize the design of SPARROW and its mesh network. We are also actively seeking partners who can integrate SPARROW Mini directly into their existing cameras for compatibility testing.

## 8.2 Alternative applications and the future of SPARROW

At its essence, SPARROW functions as a self-sustaining, centralized ecosystem for data ingestion, edge-AI preprocessing, and secure uplink transmission, purpose-built to thrive in the challenging conditions of remote and harsh environments where power intermittency, extreme weather, and logistical isolation conspire against traditional monitoring. SPARROW system's design is inherently agnostic to sensor modalities. Any data collection apparatus capable of exporting payloads via serial protocols, file dumps, or even rudimentary wireless beacons can plug into SPARROW's pipeline. The XBee mesh network with Sparrow Mini will support larger-scale deployments by enabling long-range, low-power communication across distributed nodes.

This adaptability opens opportunities for broader uses. Potential applications include:

- Disaster responses such as early warning systems for floods and fires (Cratere 2025; Lee et al. 2025; Vázquez, Zhai, and Yang 2025). The platform currently operates an on-device fire detection model, providing real-time wildfire detection and alerting in remote areas without reliance on wired infrastructure.
- Subsurface or partially submerged aquatic fauna detection and monitoring (e.g., turtles in ponds and wetlands) with specialized sensors like thermal cameras (Sellés-Ríos et al. 2022; Snyder 2017; Wilson et al. 2025).
- Microclimate and soil monitoring using arrays of sensors and small stations to track local and remote conditions.
- Gunshot detection in remote or urban settings via acoustic analysis.
- Animal tracking with telemetry devices integrated into the mesh network.
- Marine monitoring buoys, drones, or gliders for tasks such as mammal detection, seafloor mapping, and fishery assessment.

In the longer term, SPARROW aims to support an Internet of Things approach in environmental science and biodiversity monitoring, facilitating data connections between ecosystems and users on a larger scale than previously feasible. Through open-source releases and community contributions, it can help expand access to environmental data for researchers, managers, and local stakeholders.

# 9. Open Source and Code Availability

A key contribution of this work is the open sourcing of SPARROW's hardware and software stack, ensuring transparency, reproducibility, and community-driven innovation.

All necessary resources are publicly accessible via https://github.com/microsoft/sparrow-client, under a permissive license that encourages adaptation for diverse research and conservation contexts.

By making the code and documentation openly available, we aim to lower barriers for field deployment, foster collaborative improvements, and accelerate the adoption of edge AI solutions for environmental monitoring and data acquisition speed.

## 10. Acknowledgements

We thank the organizations whose collaboration made this research possible. SPARROW deployments in Colombia were carried out with the support of SINCHI, Fundación Biodiversa Colombia, Universidad de los Andes, and CINFONIA. We extend our gratitude to Instituto Humboldt, who contributed to the bioacoustics model development. In Perú, Planet and Los Amigos Biological Station of Conservación Amazonica – ACCA enabled us to deploy our system in the Amazon Rainforest. In the United States, Aaron Huey and Devlin Gandy, the Turtle Conservancy, Woodland Park Zoo, Alert California, and the University of California San Diego assisted with deployments in several regions.
We gratefully acknowledge the support provided by the Jane Goodall team in Gombe and the Salonga National Park and Virunga National Park, whose collaboration was instrumental in facilitating our deployments across African field sites.

We appreciate and are thankful for the involvement of each institution in this project.

## 12. SPARROW Assembly & Setup Instructions

*The most current version of the SPARROW Assembly & Setup Instruction Manual can be found at https://aka.ms/sparrowassembly*

## 13. SPARROW Bill of Materials (BOM)

*The most current version of the SPARROW Bill of Materials (BOM) can be found at https://aka.ms/sparrowbom*